# A closed quantum system giving ergodicity

J.M. Deutsch

Department of Physics,

University of California,

Santa Cruz CA 95064

and

The James Franck Institute

5640 South Ellis Avenue

Chicago IL 60637

A closed quantum mechanical system does not necessarily give time averages in accordance with the microcanonical distribution. This question is investigated if the number of degrees of freedom N is large. For systems where the different degrees of freedom are uncoupled, experimental situations are discussed that show a violation of the usual statistical mechanical rules. It is shown that by applying a finite but very small perturbation to such systems, the results of quantum statistical mechanics can indeed be recovered. The form of the perturbation is that of a banded random matrix, which has been used previously to describe strongly chaotic systems in the semiclassical limit. The properties of energy eigenfunctions for this perturbed system are also discussed, and deviations from the microcanonical result are shown to become exponentially small in the limit of large N.

# 1. Introduction

Justification for the Gibbs formula in computing time averaged quantities, has been discussed at great length for classical systems [1,2]. The word justification is used here rather than derivation, as at the present time there are only a small number of systems where a proof of ergodicity and mixing has been made. There are two paths that have been followed to explain, or justify why the rules of statistical mechanics appear to work. The more convincing explanation considers a complete system in isolation. From the assumption of ergodicity and mixing the above rules can indeed be derived. This can be done for a closed system completely decoupled from the external world. The second less convincing argument couples the system of interest to a heat bath. Assumptions of transfer of energy to and from the heat bath have to be made, and essentially this kind of argument pushes back the problem to another stage; justifying the statistical properties of the heat bath.

The classical justification is far more convincing than the quantum mechanical analog which has relied on several approaches. The question of the validity of the microcanonical distribution for a closed system quantum mechanical system is still regarded as an open question [2]. Von Neumann's approach [3] to this problem introduces the idea of ''macro-observables''. The actual observation is regarded here as important in understanding this question. When measurements are made, they are made by a special class of operators that allow the simultaneous measurement of any quantity of physical interest. His approach and a variety of related work has come under a great deal of criticism as it implies that all macroscopic systems are ergodic; a statement that is hard to defend with any reasonable definition of the word "ergodic" [4]. A variety of other formal approaches based on the idea of macro-observables has lead to a conclusion less general [2], however the criteria for ergodicity that have been found are not known to be satisfied by any physical systems. Much work has been done on systems that are not in isolation but coupled to an external environment [5], or subject to random perturbations, this has often emphasized the need for a coupling to the external environment in order to obtain ergodicity [2].

This paper is an attempt to justify for a closed quantum mechanical system why the laws of quantum statistical mechanics work. The philosophy taken here is to see what can be derived about statistical mechanics from quantum mechanics without any additional assumptions and for a particular choice of model systems. The assumption made in quantum statistical mechanics is that the average over time $< \cdots >_t$ of some observable quantity $< \psi | A | \psi >$ is equal to a microcanonical average at an energy a total energy $e$ that is assumed to be well defined (see below).

$$<< \psi | A | \psi >>_t = \sum_j \Delta(e, e_j) < j | A | j > \qquad (1.1)$$

where $j$ labels an energy eigenstate of the entire system and $\Delta(e, e_j)$ is a function that is sharply peaked at $e = e_j$. For a system containing a large number of degrees of freedom and for a large class of operators $A$, this can be written with negligible error in terms of the canonical distribution at fixed temperature

$$<< \psi | A | \psi >>_t = \frac{\sum_j e^{-e_j/T} < j | A | j >}{\sum_j e^{-e_j/T}} \qquad (1.2)$$

We will postpone to the end of the introduction how these formulas should, rather simply, be modified to take into account fluctuations the total energy, but (1.1) or equivalently (1.2) is has an been enormously successful in explaining problems in almost every branch of physics. For the purposes of this paper, Systems obeying the above equation will be called "ergodic".

In classical mechanics, a system with a few of degrees of freedom such as Sinai billiards, have time averages given by the microcanonical distribution. A quantum mechanical treatment of the same system cannot be expected to give the microcanonical distribution. It is easy to show by counterexample, that one needs at least one more requirement; the number of degrees of the system must also be large. Indeed, if the spacing between energy levels is not small it is impossible to define a microcanonical distribution in a precise way. Having a large number of degrees of freedom however is not enough to insure ergodicity. As a simple example, consider a perfect harmonic crystal. If we consider the expectation value of the mean square displacement of one atom, it will depend on the initial wavefunction chosen. If we restrict our attention to energy eigenstates, then for large N, almost all eigenstates will give the microcanonical answer for $< x^2 >$. Thus if all eigenstates around one energy are weighted with equal probability, we have with almost certainty that if one of these $|\psi>$ is chosen, then $<\psi|x^2|\psi> = < x^2 >_{micro}$. This is shown in Appendix C. However if the system is perturbed, say by shining light on the crystal to give it a different total energy, the distribution will now be biased by the fact that the light will couple unequally to different modes. In this case $< x^2 >$ is different than what one would get from the microcanonical distribution (this can be seen from formula (A6) and is discussed in more detail in section 5). The same behavior is also observed classically, and is a consequence of the integrability of the system. In contrast, If the crystal was not harmonic but sufficiently nonlinear, then after an external perturbation was added, the system would again give the microcanonical distribution but now at a higher energy.

The task then is to find some quantum mechanical system that does give back the microcanonical distribution after it has been perturbed as described above. The approach taken will be similar to understanding the statistical mechanics of an "ideal gas". A genuine ideal gas has no interaction between different particles and therefore will not be ergodic, but can be made so by slight modification. For example the particles can be given hard cores of very small diameter, which will have a negligible effect on the statistical and thermodynamic properties computed from the Gibbs distribution. However after a long enough time, the system will explore all of its available phase space enabling the rigorous application of these formulae. In the same spirit, suppose we start with a

Hamiltonian that decouples into N separate subsystems

$$H_0 = \sum_{i=1}^{N} h_0(x_i, p_i) \tag{1.3}$$

This system is not ergodic. However the main result of this paper, is that with negligible error in the limit of large N, it can be made ergodic by the addition of a small perturbation. The perturbation added to the Hamiltonian is a real symmetric random matrix, with certain physically sensible conditions on the statistics of the elements discussed in detail below.

The differences between the results obtained for the above system and those of an ergodic system vanish in the limit of large N, except for very special initial conditions. Thus this system provides a justification for statistical mechanics without invoking any coupling to the external world. This is interesting because as mentioned above, such a coupling has been thought necessary to give ergodicity.

The choice of a random matrix as the perturbation added to $H_0$, can be given some justification. Recently, much attention has been given to the idea that random matrices fit the spectrum of semi-classical chaotic systems, such as Sinai billiards [6]. Much numerical evidence has been accumulated to support this view point. In this work however, we are not interested in the semiclassical limit, but in large N. Numerical studies of the same type for large N have not been performed, as they are not feasible, however earlier experimental work on nuclei strongly support the conjecture, as first proposed by Wigner [7], that the spectrum of a random matrix accurately models the levels in nuclei [8]. This is important in that these nuclei are not in the semiclassical regime.

Of course the actual wave functions obtained from a random matrix model may not be such a good description, and perhaps additional physics related to locality of interactions and conservation laws should be incorporated. There is numerical and analytic work in the semi-classical limit, suggesting that this description is a good approximation [9,10].

The model system described here is simple enough to be analytically tractable, and still give non-trivial results.

Before proceeding any further, it is useful to discuss how the microcanonical formula (1.1) should be modified to take into account fluctuations in the total energy. If we have an arbitrary initial state

$$|\psi> = \sum_i \Gamma_i |i>$$ (1.4)

one could ask if it is possible to experimentally prepare a system having a very broad distribution for $\Gamma_i$, that is large quantum mechanical fluctuations in the total energy of the system. The answer appears to be that it is indeed possible.

One needs simply to superpose the wave functions of two systems, one at a low energy, and another at high energy. This can even be done practically by a contraption similar but less fiendish than devised by Schroedinger's for his fictitious cat. In this well known example, a certain apparatus was placed in a safe. It consisted of radioactive nuclei that if decayed, would trigger a detector, that would detonate a bomb that would kill a cat. If one instead were to substitute say, a brick close to zero degrees in place of the cat, then by the same reasoning as Schroedinger's the final state state of the brick should be in the superposition of two states, one hot, and one cold brick. This would give a very broad distribution in energy for $\Gamma_v$.

For a macroscopic system, one would expect that the time average of an observable could be described by the weighted average of two microcanonical averages. The weights being the probability that the brick was to be found in one or the other macroscopic states, and the microcanonical averages taken at the appropriate energies. The assumption being made is that there is negligible interference between two macroscopically different states. If we denote $<A>_e$ as the microcanonical average of an observable in a state with a well defined total energy $e$, then in general we expect the time average of $A$ for a wavefunction with a broad distribution of total energies to be

$$< A >_t = \sum_e P(e) < A >_e \qquad (1.5)$$

where P(e) is the probability of finding that the system has an energy e. In terms of $|\psi>$ as defined above we expect for an ergodic system

$$\ll \psi |A| \psi \gg_t = \sum_\nu \Gamma_\nu{}^2 < A >_{e_\nu} \qquad (1.6)$$

In the above example, a large uncertainty in the total energy is obtained by the addition of an external system, or potential, whose value is uncertain. We can ask the question as to what happens if a known external potential is applied to the system and then switched off. For simplicity let us consider the time evolution of the integrable case. We add an external potential of the form

$$\sum_{i=1}^{N} V(x_i, p_i, t) \qquad (1.7)$$

to $H_0$. Then if the system starts out in an eigenstate of $H_0$, we can ask what is the spread in the total energy

$$\Delta E^2 = \sum_\nu \nu^2 \Gamma_\nu < \psi |(H- < \psi |H| \psi >)^2| \psi > \qquad (1.8)$$

after the external perturbation has been turned off. An example of such a system would be to shine light momentarily on a harmonic crystal. Since the Hamiltonian separates, the different degrees of freedom evolve separately, so that

$$< \psi |(H- < \psi |H| \psi >)^2| \psi > = \sum_i < \psi |(h_i- < \psi |h_i| \psi >)^2| \psi > \qquad (1.9)$$

That is, the total variance in the energy is the sum of the variances of the individual particles. Therefore it is expected that the $\Delta E/E$ will decrease as $1/N$. Therefore applying an external potential to this system, which initially has a well defined energy, will not create macroscopic fluctuations in its state, that is $\Gamma_\nu$ remains localized around a small range of energies.

The organization of this paper is the following. In section 2, the model Hamiltonian that will be used is defined and a heuristic discussion of what it is expected to yield is briefly given. In section 3, the statistical properties of the eigenvectors of this Hamiltonian are derived. Using these results, various quantities of interest can be computed. In section 4, properties of the system in energy eigenstates are analyzed and compared to unperturbed, that is ''integrable'' eigenstates. Section 5 deals with the time averages of the system that started in an eigenstate of the integrable system. The most general initial condition is considered in section 7, where it is shown that time averages give the micro-canonical result (1.6) with almost certainty in the limit of large N. These results are discussed in the context of some *gedanken* experiments in section 7, which contrast results experiments done on ''integrable'' systems to systems giving thermodynamic equilibrium, such as the model described here. Finally further questions of interest using this model are presented.

## 2. The Model

We will consider the real symmetric Hamiltonian

$$H = H_0 + H_1 \tag{2.1}$$

$H_0$ could represent say, the Hamiltonian of a harmonic crystal, or describe an ideal gas. In this work we shall be interested in the limit of a large number of degrees of freedom, and shall see that considerable simplifications take place in this limit. $H_1$ is added in the hopes of making the system ergodic. In the case of an ideal gas for example, one may want to add some interaction between the different particles, for example take

$$H_1 = \sum_{i<j=1}^{N} V(r_i - r_j) \tag{2.2}$$

Instead of adding in these interactions explicitly, we model $H_r$ by a real symmetric matrix whose elements are chosen from a random Gaussian ensemble.

It will be convenient to examine this model in the basis where $H_0$ is diagonal, that is the basis of eigenvectors of $H_0$. In addition, we can see that matrix elements of $H$ between two different energies $< E_1|H_1|E_2 >$ must decrease on average as $E_1 - E_2$ increases when $H_1$ is given by (2.2). The size of elements are diminished by a phase space factor

$$< E_1|H_1|E_2 > \tilde{}\ e^{|E_1 - E_2|/T} \qquad (2.3)$$

for $T \ll |E_1 - E_2| \ll E_1$ (the ground state energy of the system is set to 0). The temperature is defined as usual as

$$\frac{1}{T} = \frac{\partial S}{\partial E} \qquad (2.4)$$

A justification for this is given in Appendix B. The energy $E$ in the above derivative can be evaluated at either $E_1$ or $E_2$ since for a large system $T \ll E$. So when $|E_1 - E_2| \gg T$, the proportion of non-zero matrix elements is effectively zero.

So we will model $H_1$ as a random symmetric matrix

$$h_{ij} \equiv < E_i|H_1|E_j > ,\, < h_{ij}h_{kl} > = \varepsilon^2 \delta ik \delta jl \qquad (2.5)$$

keeping in mind that the elements must be cutoff for $|E_1 - E_2| \gg T$. For finite $T$ and large N this cutoff makes no difference to many of the calculations to be presented, except at some points in Sections 4,5, and 6 where it prevents an unphysical divergence in expectation values as can be seen in more detail in Appendix A. For small N, the width of the matrix $H_1$ have been discussed [10].

It is emphasized that no ensemble average is being taken. The elements of $H_1$ are precisely determined, but have correlations between them typical of a random matrix. The same philosophy has been adopted by others in studying the semi-classical limit of chaotic systems [6].

We now discuss what we intuitively expect this model to yield. For small enough $\varepsilon$ eigenvectors will differ only slightly from their $\varepsilon = 0$ value. However for small but finite $\varepsilon$ many neighboring levels $\Delta E$ will be coupled. This is because the the distance between neighboring levels [11] $D(E) \propto exp(-S(E))$, where $S(E)$ is the total system entropy at energy E. So at fixed energy per particle, the separation between levels decreases exponentially with N and for large enough N can therefore be made arbitrarily small. So one expects that there is a large range of values for $\varepsilon$ which will couple a large number of levels, the number is $\propto \Delta E \exp(S(E))$ We would expect that for large N, $\varepsilon$ could be made much smaller than the energy per particle, and have a large effect on eigenvectors. The new eigenvectors should then mix together with random phases, the unperturbed eigenvectors within a window $\Delta E$. It is this mixing that makes the system ergodic.

In the next section we shall analyze quantitatively the distribution of eigenvectors. With this information we will be able to examine physical properties of this model in eigenstates of the Hamiltonian, and also compute time averages of observables.

## 3. The probability distribution of the eigenvectors

Given the Hamiltonian (2.1) we first would like to compute the probability distribution of the eigenvectors given that $H_1$ is drawn from a Gaussian random ensemble as described above. We can rewrite (2.1) in matrix form

$$H_{ij} = f_i \delta_{ij} + h_{ij}, \ f_i = <E_i|H_0|E_i> .$$ (3.1)

There appears to have been little work on this kind of random matrix problem. The change in the density of states has been considered previously [12], but not the structure of the eigenvectors. The effect on the density of states in the limit we are interested in, is minor. It has the effect of reducing fluctuations and in particular gets rid of degeneracies in the $\varepsilon = 0$ Hamiltonian. Given a particular matrix $H$, we are interested in the matrix $c$ satisfying

$$c^T H c = \Lambda \tag{3.2}$$

where $\Lambda$ is a diagonal matrix. We will add the constraint that $c$ be normalized, that is $cc^T = I$. Since the above equation is linear in $H$, we can write the probability distribution for the matrix c as

$$P(c) \propto \delta(cc^T - I) \int e^{(\sum_{i \neq j} H_{ij}^2 - \sum_i (H_{ii} - f_i)^2)/(2\varepsilon^2)} \prod_{i \neq j} \delta(\sum_{kl} c_{ki} H_{kl} c_{lj}) \prod_{ij} dH_{ij}. \tag{3.3}$$

Using the Fourier representation of the $\delta$ function these integrals can be carried out giving

$$P(c) \propto \delta(cc^T - I) e^{\sum_i (\sum_k f_k c_{ki}^2)^2/(2\varepsilon^2)} \tag{3.4}$$

As a check of this result, note that when $H_0$ is a constant, this distribution becomes uniform.

Next we would like to find the probability distribution of one eigenvector, that is, integrate out over all $c_{ij}$ except for one i. It has not been possible to do so exactly, but an approximate method has been found that should work well in the limit we are interested in. The idea is as follows. We are interested in the statistical properties of (3.4). We can think of this expression as representing N vectors that are restricted to be orthonormal and weighted by the above exponential. We are considering a regime where the eigenvectors are strongly overlapping. Consider one of these vectors and label its components $a_i$. The sign of the different $a_i$ should shift back and forth, to insure orthogonality, but we expect the overall magnitude of the components should show a maximum in some range of i. In the region of strong overlap we therefore expect to be able to characterize the probability distribution of the $a_i$'s by writing

$$P\{a_i\} \propto e^{-\sum \frac{a_i^2}{2\Lambda_a(i)}} \tag{3.5}$$

$$\Lambda_a(i) = \langle a_i^2 \rangle \tag{3.6}$$

We are interested in finding the functional form of $\Lambda_a(i)$ for each eigenvector. We shall take as a trial distribution, the product of (3.5) over all eigenvectors, and substitute it into (3.4) integrating over all the $c_{ij}$'s. The form of $\Lambda$ giving the largest result will give an estimate of $\langle a_i^2 \rangle$. In doing this maximization, we have to correctly take into account the "entropy" associated with each $\Lambda$, that is the number of different configuration of the $a_i$'s that are associated with each $\Lambda$. This procedure amounts to applying a standard variational principle of statistical mechanics [13].

The restriction of orthogonality as given in (3.4) gives rise to an effective interaction between two eigenvectors a and b that can be obtained by integrating over the $c_i$'s subject to the orthonormality constraint. Thus we consider

$$Z = \mathop{tr}_{a,b} P\{a_i\} P\{b_i\} \delta(|a|^2 - 1)\delta(|b|^2 - 1)\delta(a \cdot b) \tag{3.7}$$

The interaction between vectors a and b comes through the orthogonality restriction of the last $\delta$ function. Without this the partition decouples and with the correct normalization one obtains $Z = 1$. With orthogonality enforced, Z can be evaluated by using the Fourier representation of the $\delta$ function constraints

$$Z = \mathop{tr}_{a,b} \int d\eta e^{\eta} \int d\lambda \int d\nu e^{\nu} e^{-\sum_i \frac{a_i^2}{2\Lambda_a(i)} - \frac{b_i^2}{2\Lambda_b(i)} - \eta a_i^2 - \nu b_i^2 - \lambda a_i b_i} \tag{3.8}$$

Performing the traces over a and b give

$$Z \propto \int \exp^{-\frac{1}{2}\sum_i \ln\left[(\eta + 1/\Lambda_a(i))(\nu + 1/\Lambda_b(i)) - \lambda^2\right]} d\lambda d\nu d\eta \tag{3.9}$$

Locating the saddle point of $\eta$, $\nu$ and $\lambda$ gives $\lambda = \nu = \eta = 0$. Expanding the logarithm to second order in $\lambda$, we obtain

$$Z \propto \left( \sum_i \Lambda_a(i) \Lambda_b(i) \right)^{-1/2} \qquad (3.10)$$

In the limit of a broad distribution for $\Lambda$, higher order terms in this Taylor series have a negligible effect. Since we are interested in the limit of strongly overlapping eigenvectors, we are justified in only keeping the lowest order term.

Now we are in a position to formulate a ''free energy'' as a function of $\Lambda$ and minimize to obtain our final solution. We are no longer interested in the interaction between just two eigenvectors but the interactions among all n, so we modify our notation writing

$$< c_{ij}^2 > = \Lambda(i, j) \qquad (3.11)$$

The entropy associated with a particular collection of $\Lambda$'s is

$$S = \frac{1}{2} \sum_{ij} \ln \Lambda(i, j) \qquad (3.12)$$

The inter-vector repulsion is obtained by summing over all pairs of vectors

$$E_{rep} = \frac{1}{4} \sum_{ij} \ln(\sum_m \Lambda(m, i) \Lambda(m, j)) \qquad (3.13)$$

which can be obtained less heuristically by modifying (3.7) to constrain the n eigenvectors $c_{ij}$ to be orthonormal

$$Z = \frac{tr}{c} \, e^{-\sum_{ij} \frac{c_{ij}^2}{2\Lambda(i,j)}} \prod_l \delta(\sum_k c_{kl}^2 - 1) \prod_{i<j} \delta(\sum_k c_{ki} c_{kj}) \qquad (3.14)$$

and computing by the same technique as above. The distribution (3.4) is biased by the exponential factor, giving

$$E_b = \frac{-1}{2\varepsilon^2} \sum_i \left( \sum_k f_k c_{ki}^2 \right)^2 \tag{3.15}$$

When $\Lambda(i, k)$ is sufficiently slowly varying the above expression is self averaging and can be replaced by

$$E_b = \frac{-1}{2\varepsilon^2} \sum_i \left( \sum_k f_k \Lambda(i, k) \right)^2 \tag{3.16}$$

So we can write the total ''free energy'' to be minimized as

$$F = E_{rep} + E_b - S \tag{3.17}$$

We would like to minimize F with the constraint

$$\sum_i \Lambda(i, j) = \sum_j \Lambda(i, j) = 1 \tag{3.18}$$

(which is obtained by averaging $\sum_i c_{ij}^2 = \sum_j c_{ij}^2 = 1$). Therefore we wish to solve

$$\frac{\partial F}{\partial \Lambda(\mu, \nu)} + \lambda_\mu + \kappa_\nu = 0 \tag{3.19}$$

where $\lambda$ and $\kappa$ are Lagrange multiplier coefficients. Also we are interested in the regime where the width of $\Lambda$ is large, but still small compared with the energy per particle, thus we can write

$$f_k = \Delta\, k \tag{3.20}$$

where $\Delta$ is the average spacing between levels. Therefore

$$\frac{\partial F}{\partial \Lambda(\nu, \mu)} = -\frac{a^2}{\varepsilon^2} \nu \sum_k \Lambda(k, \mu) k + \frac{1}{2} \sum_i \frac{\Lambda(\nu, i)}{\sum_m \Lambda(m, i) \Lambda(m, \mu)} - \frac{1}{2\Lambda(\nu, \mu)} \tag{3.21}$$

We also expect that for eigenvectors whose eigenvalues do not lie near the extremities,

$\Lambda(i, j) = \Lambda(i - j)$. This is because the condition $\sum\limits_{j} \Lambda(i, j) = 1$ means the system of eigenvectors are "incompressible", that is the eigen vectors cannot bunch up in any one energy value. This choice for $\Lambda(i, j)$ also satisfies (3.18). The first term in (3.21) simplifies since

$$\sum_{k} \Lambda(k, \mu)k = \sum_{k} \Lambda(k - \mu)k = \mu \qquad (3.22)$$

(here we can choose $\sum\limits_{k} \Lambda(k) = 0$ by a trivial shift in k.

To solve (3.21) we guess that the solution is of the form of a Lorentzian

$$\Lambda(i, j) = \frac{A}{(i - j)^2 + \delta^2} \qquad (3.23)$$

and surprisingly find that it does indeed work. Since we are interested in $\delta$ large, the summations in (3.21) can be replaced by integrals, and the integrals can be carried out. Carrying out the integration and choosing the $\lambda\mu$ and $\kappa\nu$ to cancel unwanted terms, we obtain

$$A = \frac{\varepsilon^2}{2\Delta^2}, \quad \delta = \frac{\pi\varepsilon^2}{2\Delta^2} \qquad (3.24)$$

This result can be compared with the opposite limit of $\varepsilon \ll \Delta$, one can easily show that to order $\varepsilon^2$

$$< c_{ij}{}^2 > = \frac{\varepsilon^2}{\Delta^2} \frac{1}{(i - j)^2} \qquad (3.25)$$

For large $i - j$ these two limits differ only by a factor of 2.

In summary, we have obtained an approximate form for the probability distribution of one eigenvector. After deriving the complete probability distribution of all eigenvectors, it was necessary to integrate over all but one of the vectors. In deriving the distribution of one eigenvector, it was useful to think of this situation as a statistical mechanical problem involving the interaction between N eigenvectors that all can have arbitrary distributions $\Lambda(i, j)$. For fixed i, $\Lambda(i, j)$ can be thought of as the density of some fictitious macromolecule along a one dimensional space j, with a constant mass (fig. 1). The restriction of normality implies incompressibility. Thus the different $\Lambda(i, j)$'s can lie on top of each other must have a constant total density. Therefore we expect that the system will look translationally invariant, $\Lambda(i, j) = \Lambda(i - j)$. Their are several competing effects giving the final distribution for $\Lambda$. There is an entropic term that favors each $\Lambda$ being spread out. There is a bias which favors a narrow distribution. Finally the effect of orthogonality between eigenvectors gives rise to a repulsion between them. This also favors a narrow distribution. The resulting form for $\Lambda$ can then be found, which is a Lorentzian with width given by (3.24).

As a further check on this approach, if the effect of intervector repulsion (3.13) is ignored, but not the incompressibility constraint, the problem can be solved exactly, and the variational calculation gives the same answer, which is a Lorentzian with twice the width obtained above in (3.24). Thus the inclusion of the orthogonality constraint reduces the width of the distribution obtained by a factor of two.

## 4. Observables in Eigenstates

Before discussing time averaged properties of observables, it is illustrative to consider the properties of observables in energy eigenstates. This will highlight the difference between a non interacting system and one with very small but finite $\varepsilon$. We will see that fluctuations in results are exponentially reduced by the presence of a very small interaction of the form (2.5).

We denote the state vector of the interacting Hamiltonian in the ith energy eigenstate by $|i>$. A noninteracting state vector in the jth energy eigenstate is labeled $|j>_0$. They are related by

$$|i> = \sum_j c_{ij} \, |j>_0 \qquad (4.1)$$

We want to look at the variation in the expectation value of an operator A in an energy eigenstate

$$< i|A|i > = \sum_{kl} c_{ik} c_{il} \,_0 < k|A|l >_0 \qquad (4.2)$$

We will restrict our attention to real A. Since energy eigenstates for neighboring energy levels look different, it is possible for $< i|A|i >$ to vary from state to state. We are interested in calculating the variation of $< i|A|i >$ in a microscopic energy range. If we compare the expectation value at two energies that differ by more than $\delta$ but still much less than a macroscopic energy, then the two eigenvectors become completely uncorrelated. Therefore to compute the variance in $< i|A|i >$ between states separated by such energies, we can compute the variance in $< i|A|i >$ keeping state i fixed, but averaging different realizations of the random matrix $H_1$. Denote this kind of averaging by $< \cdots >_{rand}$. Therefore a good measure of the variance is

$$\Delta A^2 \equiv << i|A|i >^2>_{rand} - << i|A|i >>^2_{rand} \qquad (4.3)$$

First note that the second term on the right hand side of this equation involves $<< i|A|i >>_{rand}$ which can be computed by averaging (4.1) giving

$$<< i|A|i >>_{rand} = \sum_j \Lambda(i, j) \,_0< j|A|j >_0 \qquad (4.4)$$

This is just the microcanonical average of A $< A >_{micro}$ (see Appendix A). This is fortunate, as this gives the ergodic mean. The first term on the right hand side of (4.3) can be computed by averaging the square of (4.2) giving

$$<< i|A|i >^2>_{rand} = \sum_{jklm} < c_{ij} c_{ik} c_{il} c_{im} >_{rand} \,_0< j|A|k >_0 \,_0< l|A|m >_0 \qquad (4.5)$$

We shall assume as we did in the last section, that the c's are Gaussian. This can be shown explicitly when $\frac{\varepsilon}{a}$ is infinite, and should be a good approximation when such a quantity is large. Therefore

$$< c_{ij} c_{ik} c_{il} c_{im} >_{rand} = \qquad (4.6)$$

$$\Delta(i,j)\delta_{jk}\Delta(i,l)\delta_{lm} + \Delta(i,j)\delta_{jl}\Delta(i,k)\delta_{km} + \Delta(i,j)\delta_{jm}\Delta(i,k)\delta_{kl}$$

Substituting this into (4.5) gives

$$\Delta A^2 = << i|A|i >^2 >_{rand} - \left( \sum_j \Lambda(i,j) \, _0< j|A|j >_0 \right)^2 = 2 \sum_{jk} \Lambda(i,j)\Lambda(i,k) \, _0< j|A|k >_0^2$$

$$\leq 2 \sum_{jk} \Lambda(i,j)\Lambda(i,i) \, _0< j|A|k >_0^2 = 2\Lambda(i,i) \sum_j \Lambda(i,j) \, _0< j|A^2|j >_0 \qquad (4.7)$$

$$= 2\Lambda(i,i) \, _0< i|A^2|i >_0 >_{rand} = 2\Lambda(i,i) < A^2 >_{micro}$$

From (3.23) and (3.24) we have that $\Lambda(i,i) = 1/(\pi \delta)$. We will keep the spread in energy of the eigenvectors $\Delta\delta$ constant as the number of degrees of freedom $N$ goes to $\infty$. But because at a fixed energy $\Delta \propto \exp(-const \times N)$, $\Lambda(i,i) \propto \exp(const \times N)$. For a large class of operators, the ones related to extensive and intensive statistical quantities (.e.g. total energy, correlation functions, susceptibilities) $< A^2 >_{micro}$ increases with no more than a power of N. Therefore we expect that $\Delta A^2$ should decrease exponentially with N. This should be contrasted with the case $\varepsilon = 0$ which is now examine.

If one considers fluctuations in the expectation value of A with $\varepsilon = 0$ then the variance in such a system should be

$$\Delta A^2{}_0 = < \, _0< i|A|i >_0^2 >_{micro} - << \, _0< i|A|i >_0 >^2{}_{micro} \qquad (4.8)$$

where again $< \cdots >_{micro}$ denotes a microcanonical average, one that is done over a small width in energy. We can compute the value of $\Delta A^2{}_0$ for the example of a harmonic crystal as done in Appendix C. The result is that $\Delta A^2{}_0$ decreases with a negative power of $N$. Thus for large N the fluctuations in the integrable case are dramatically larger than in the $\varepsilon \neq 0$ case.

The reason why the interacting case shows much smaller fluctuations can be intuitively understood as the result of an additional averaging done over neighboring energy states as given by (4.1). The fluctuations are then reduced by the square root of the number of states being averaged over, which in this case is $1/\Lambda(i, i)$ which is exponential in N.

It is this reduction in these fluctuations which is responsible for the agreement between ergodic results and the random matrix model described here. We next derive some results for time averaged quantities which bear a striking resemblance to those of an ergodic system.

## 5. Time averages starting from an integrable eigenstate.

We consider an integrable system with Hamiltonian of the form (1.3) in an energy eigenstate $|\psi> = |e>_0$. At time $t = 0$ an interaction of the form (3.1) is turned on. We compute the time averaged value of an operator A defined as

$$<< \psi|A|\psi >>_t \equiv \lim_{T \to \infty} \int_0^T < \psi|A|\psi > dt \qquad (5.1)$$

In terms of the interacting basis (where there should be no degeneracy), this can easily shown to be

$$<< \psi|A|\psi >>_t = \sum_i | < i|\psi > |^2 < i|A|i > \qquad (5.2)$$

This is known as the ''Fine-grained Ergodic Theorem'' [2,3]. Before proceeding, it is useful to again point out that in the integrable case, this time average does not in general give the ergodic result. Consider again the example of a harmonic crystal. In Appendix C, the mean square displacement of one atom $< \psi | x^2 | \psi >$ is computed in an energy eigenstate, eqn. (C6). It depends on the frequency of each normal mode $\Omega_\nu$, and the occupancy of each mode $n_\nu$. Two different sets of $n_\nu$ that have the same total energy, will typically also have the almost the same value for $< \psi | x^2 | \psi >$. This is because for a large system the number of modes in a given energy interval is large so that (C6) is self averaging. However one can find energy eigenstates that are very atypical. For example, one could let all the $n_\nu$'s equal zero except one which must then be very large in order to give the same total energy. What is shown below, is that even starting out with this state, the time average in the interacting case is identical to the microcanonical result, in the large N limit.

Inverting (4.1) to express the interacting state $|e>_0$ in terms of interacting ones

$$|e>_0 = \sum_j c_{je} |j>$$ (5.3)

and therefore $<< \psi | A | \psi >>_t$ can be rewritten in terms of the noninteracting set of vectors

$$<< \psi | A | \psi >>_t = \sum_{ijk} c^2{}_{i,e} c_{j,i} c_{k,i} \, _0< j | A | k >$$ (5.4)

We are interested in seeing if this quantity gives a similar result to to that obtained for an ergodic system. Since the $c_{i,j}$'s are randomly chosen, in order to compute this time average, we must do an average over different values of $c_{i,j}$. Fluctuations in this result can be calculated by similar means. Performing this average, again assuming the $c_{i,j}$ obey Gaussian statistics, one obtains

$$<<< \psi | A | \psi >>_t>_{rand} = \sum_{i,j} \Lambda(e,i) \Lambda(i,j) \, _0< j | A | j >_0 + 2 \sum_i \Lambda^2(i,e) < i | A | i >$$ (5.5)

The first summation can be simplified by noting

$$\Lambda_2(e,j) \equiv \sum_i \Lambda(e,i) \Lambda(i,j)$$ (5.6)

is still of the form (3.23) and still normalized but with twice the width $\delta = \pi(\frac{\varepsilon}{\Delta})^2$. The second summation in (5.5) is easily seen to be bounded by

$$\Lambda(e, e) \sum_i \Lambda(i, e) < i|A|i >$$ (5.7)

which is exponentially smaller than the first term and is therefore negligible. We are then left with

$$<<< \psi|A|\psi >>_t >_{rand} = \sum_j \Lambda_2(e, j)\ _0< j|A|j >_0$$ (5.8)

This represents a sum over a small energy window of width $\delta\Delta$ of the diagonal elements of $A$, which is the microcanonical distribution.

Fluctuations in this average can easily be calculated and are exponentially small for reasons similar to the ones given in the previous section. Thus if the system starts in an energy eigenstate of the integrable system its time average behavior will, with exponentially small uncertainty, be given by the microcanonical distribution. This contrast with the integrable case which can give expectation values very different from the microcanonical distribution. We will now examine time averages in the general case of an arbitrary initial wavefunction.

## 6. Time averages starting from an arbitrary state

Consider an arbitrary initial state

$$|\psi > = \sum_i \Gamma_i |i >_0$$ (6.1)

evolving under the action of the interacting Hamiltonian (2.1). Here $\Gamma_i$ is an arbitrary a complex amplitude. Then from (4.1) we can write the time average as

$$<< \psi|A|\psi >>_t = \sum_{\mu\nu} \sum_{ijk} \Gamma_\nu^* \Gamma_\mu c_{\nu,i} c_{\mu i} c_{ji} c_{ki}\ _0< j|A|k >_0$$ (6.2)

Again averaging this with respect to the distribution of $c_{ij}$ gives

$$<<< \psi|A|\psi >>_t>_{rand} = \sum_\nu |\Gamma_\nu|^2 \sum_i \Lambda_2(i,\nu) < i|A|i > \qquad (6.3)$$

$$+ 2 \sum_{\nu\mu} \Gamma_\nu{}^* \Gamma_\mu \Lambda_2(\mu,\nu) \, _0< \nu|A|\mu >_0$$

The first summation represents the superposition of many different microcanonical ensembles of energy $e_\nu$ with the weight $\Gamma_\nu$. This would be ones first guess as mentioned in the introduction. The second summation represents an interference term. It will now be shown that for a large class of observables, this term is indeed negligible in comparison with the first. We use the Schwartz inequality to write

$$\sum_{\nu\mu} \Gamma_\nu{}^* \Gamma_\mu \Lambda_2(\mu,\nu) \, _0< \nu|A|\mu >_0 \leq \sqrt{\sum_{\nu\mu} |\Gamma_\nu|^2 \Lambda^2{}_2(\mu,\nu)} \sqrt{\sum_{\nu\mu} |\Gamma_\mu|^2 \, _0< \nu|A|\mu >_0{}^2} \quad (6.4)$$

$$\leq \sqrt{\sum_\nu |\Gamma_\nu|^2 \Lambda_2(\nu,\nu) \sum_\mu \Lambda_2(\mu,\nu)} \sqrt{\sum_\mu |\Gamma_\mu|^2 \, _0< \mu|A^2|\mu >_0} = \sqrt{\Lambda(0,0)} \sqrt{\sum_\mu |\Gamma_\mu|^2 \, _0< \mu|A^2|\mu >_0}$$

$\Lambda(0)$ decreases exponentially with increasing N. The second square root is a weighted average of the diagonal elements of $A^2$. Following the same logic as in section 4 we expect that this average should increase with no more than a power law in N. Therefore the second term can be ignored, and we obtain a simple formula for the time average of the expectation value of $A$ in an arbitrary state, identical to what was expected from the introduction.

## 7. Discussion

With the aid of these new results one can discuss some *gedanken* experiments constructed to highlight the main points of this paper.

Consider a perfect harmonic crystal. If the system is in an energy eigenstate then the the time average of the quantum mechanical expectation of an observable is, almost certainly in agreement with the results for an ergodic system. To be more precise, if we pick energy eigenstates within some energy window with uniform probability, then fluctuations around the microcanonical result found decrease as a negative power of $N$. One can still find many states that do not agree give the ergodic result, however the vast majority of them do. This is discussed in Appendix C.

If we now shine a laser on this system, to bring it to a new energy, then although the quantum mechanical uncertainty in its energy is very small, time averages will not give the ergodic result. This is because the external perturbation will couple most strongly to some phonon modes, giving rise to a distribution significantly different from the Bose distribution. This can be seen from (C.6)

If this same experiment were repeated with the non integrable Hamiltonian (2.1) then one would obtain the ergodic result, with exponentially small uncertainty. Therefore the system comes to a new thermodynamic equilibrium. To understand the reasoning behind this conclusion, arrived at formally in section 6, we start by observing that the wave function can be written as an arbitrary superposition of ''integrable'' wave functions. We are now considering the case where these integrable wave functions are atypical, that is give answers different than those of an ergodic system. In terms of the eigenstates of the full Hamiltonian, such atypical wavefunctions are a superposition of many eigenstates, $\delta \propto \exp(const. \times N)$. It is the eigenstates of the full Hamiltonian that are important in determining time averages as is evident for (5.2). The coefficients involved in this superposition are random, and have nothing to do with how atypical each wavefunction is. From section 4 we know that each one of these eigenstates will, with exponentially small uncertainty, give the ergodic result. Hence by tracing over $\delta$ such states, we expect to wash out fluctuations even further, giving the microcanonical distribution.

It must be emphasized that in deriving the microcanonical distribution, no average over different eigenstates has been performed. The only kind of averaging done, has been over the random matrix $H_1$. For two different realizations of the matrix $H_1$, it is seen that the answers differ typically by a quantity exponentially small in $N$. An important feature of this model system is that nearly every energy eigenstate gives the microcanonical distribution, that is expectation values are almost exactly the same as would be obtained by an average over an energy shell. Of course for a large system, to prepare a system in an energy eigenstate is practically impossible. Conceptual however, this is an important point, and explains why the "Fine-grained Ergodic Theorem" (5.2) does in fact give results in accord with quantum statistical mechanics that is (1.6). It therefore does not appear necessary to introduce coarsed grained observables as numerous authors have done [2,3], or to couple the system to an external environment [5], in order to obtain ergodicity.

Throughout this paper, reference has been made to a Hamiltonian $H_0$ that is "integrable". This was in order to contrast the two types of behavior seen with and without the addition of a finite by very small $H_1$. In performing the derivations of formulas in sections 4,5, and 6, $H_0$ could be arbitrary. In general however we expect most "nonintegrable" Hamiltonians, for large N, to be ergodic, and the addition of a small additional perturbation $H_1$ does not change this result. Though this work emphasizes systems with large N, the model may be applicable to any system where the density of states is large, say in the semi-classical limit of a system involving few degrees of freedom [7].

The work presented in section 3 predicts a Lorentzian distribution for the overlap between integrable states and those of the perturbed Hamiltonian. This could be tested numerically in the semi-classical limit. For example, a Sinai billiard with small diameter, could be compared with the unperturbed case of a free particle on a torus. It would also be interesting to test some predictions of this work for large N. This is rather difficult numerically but may be feasible presently for a spin systems of up to 10 spins. A salient feature of this work is that one eigenstate should give averages in accord with the micro-canonical distribution.

The model presented here has interesting applications to the problem of quantum dissipation [14]. The problem considered is how to introduce dissipative effects in quantum mechanics. The most successful approach has been to couple a particle to an infinite number of linear degrees of freedom. The model used here is in some sense, far more nonlinear and may be more realistic in the case of systems that are classically nonlinear, such as a gas at low temperatures. The dynamics of this system are being investigated presently.

**Acknowledgments**

I thank Dr. A.P. Young, and Mario Feingold for useful discussions, and wish to especially thank Dr. Angelo Barbieri for many useful comments and discussions. This work was supported by NSF grant DMR 87-21673.

**Appendix A**

Considering a system with N degrees of freedom $N \gg 1$, the density of states $\rho(E)$ is related to the entropy per particle $s(E/N)$ by

$$\rho(E)dE = Ce^{Ns(E/N))}dE \qquad (A1)$$

The average spacing between levels is $\Delta = 1/\rho(E)$. We are interested in estimating integrals of the form

$$\int_0^\infty \rho(E)\Lambda_2(E-e)A(E) \qquad (A2)$$

A(E) is averaged over some energy window $\ll \delta$, and will not vary significantly over the scale of energies we will be considering below namely $|E - e| < T$. $\Lambda_2(E)$ is a Lorentzian of the form

$$\Lambda_2(E) = \frac{2\delta/\pi}{E^2 + 4\delta^2} \qquad (A3)$$

where $\delta$ was defined in (3.24). However because elements $< E|H_r|E' >$ of the random matrix are cut off when $|E - E'| > T$ and for simplicity have been set identically to zero, the $c_{ij}$ will decay to zero *faster than an exponential* for $\Delta|i - j| \gg T$. We argue this as follows. If one replaces the nonzero elements of $H_r$ all by a nonzero constant, then states should become less localized, as all disorder has been eliminated (this represents a one dimensional tight binding model). In this case the states are still more than exponentially localized falling off asymptotically [15] as $|i - j|^{const.|i-j|}$. This localization is entirely due to the increase in the elements of $H_0$. Therefore when random matrix $H_r$ is added, we expect even stronger localization. Therefore $\Lambda_2(E)$ is cutoff for $|E| > T$.

We expand $\rho(E)$ about $e = E$. Using (A1) we have

$$\rho(E) = \rho(e)e^{(E-e)/T} \tag{A4}$$

substituting this into the integrand of (A2), the maximum is shifted from 0 to $(E - e) = 2\delta^2/T$. A minimum now develops at $(E - e) = T/2$. With an upper cutoff on the integration of $E_{max}$, the above integrand (A2) integrated between $T/2$ and $E_{max}$ is bounded by

$$\frac{8}{\pi}\delta e^{(E_{max}-e)/T} \tag{A5}$$

With a cutoff placed on $(E_{max} - e)$ at a contant times $T$, this is negligibly small in the limit of large N.

## Appendix B

In this appendix we justify in more detail the cutoff on the random matrix used in calculations. We are interested in determining how $< E|V|E' >$ varies with increasing $E - E'$. The quantity we wish to compute is $<< E|V|E' >>_{E,E'}$ Here the second set of brackets denotes a microcanonical average over both $E$ and $E'$. $V$ is taken to be of the form (3), and we consider a system of identical particles that are either fermions or bosons. Label the eigenstates of a single particle by i. The energy in that state is labeled $e_i$, the total number of particles in state i is $N_i$ and $N_i \leq 1$ for the case of fermions. In second quantized notation the total wave-function can be written as

$$|E> = \prod_i \frac{a^{\dagger}{}_i{}^{N_i}}{\sqrt{N!}}|0>$$

where $|0>$ is the ground state. The potential $V$ can also be written in second quantized form as

$$V = \sum_{jklm} V_{jklm} a^\dagger_j a^\dagger_k a_l a_m$$

$$V_{jklm} = \int \psi^*_j(r) \psi^*_k(r') V(r-r') \psi_l(r) \psi_m(r') dr dr'$$

Thus the quantity we wish to compute is

$$<< E|V|E' >>_{E,E'} = \frac{1}{N(E)\ N(E')} \sum_{N_i's} \sum_{N_j's} \delta(E - \sum_i N_i e_i) \delta(E' - \sum_i N_i' e_i') < E|V|E' >$$

where $N(E)$ is the appropriate normalization. In the above, $\sum_{N_i's}$ means the sum over all possible combinations of $N_i's$ with the constraint $\sum_i N_i = N$. Writing the above equation in second quantized form and taking the inner products gives

$$<< E|V|E' >>_{E,E'} = \frac{1}{N(E)\ N(E')} \sum_{jklm} \sum_{N_l',N_m'} \sum_{N_i's} \delta(E - \sum_i N_i e_i) \delta(E - E' - (e_j + e_k - e_l - e_m)) V_{jklm}$$

$$= \frac{1}{N(E')} \sum_{jklm} \sum_{N_l',N_m'} < \delta(E - E' - (e_j + e_k - e_l - e_m)) \sqrt{N_j N_k N_l' N_m'} V_{jklm} >_E$$

where the last bracket denotes a microcanonical average at energy $E$. As long as $|E - E'| \ll E$, and N is large, the microcanonical average here can be replaced by a canonical average at the appropriate temperature T as given in (2.4) where

$$S = \log(\sum_{N_i's} \delta(E - \sum_i N_i e_i))$$

If we consider potentials $V(r)$ which have a Fourier transform that is bounded, then so is $V_{jklm}$ as the single particle eigen states are plane waves. We can therefore bound the above equation by

$$\frac{1}{N(E')} \sum_{jklm} \sum_{N_l',N_m'} < \delta(E - E' - (e_j + e_k - e_l - e_m)) \sqrt{N_j N_k N_l' N_m'} >_T$$

Now consider what happens for $E - E' \gg T$. In this limit it is straightforward to substitute in the appropriate bose or fermi distributions for each $N_i$ and perform the summations, but the asymptotic result can be seen by the following argument. The above average only has contributions to it when $e_j + e_k = e_l + e_m + E - E'$. If $E'$ is kept fixed and E is increased then the minimum energy needed to obtain an contribution occurs when $e_l = e_m = 0$, so that $e_j + e_k = E - E'$.(Here we are setting the ground state energies equal to zero.) As E is increased the weight of having such a configuration is given by the appropriate bose or fermi distributions which asymptotically give a weight of $\exp(-e_j - e_k) = \exp(-(E - E')/T)$. Considering larger $e_l$ and $e_m$ does not change the above exponential dependence, but just the overall prefactor.

## Appendix C

In this appendix, some properties of a quantum harmonic crystal will be analyzed. We consider a crystal in d dimensions containing N lattice sites, that forms a cube with sides of length $\propto N^{1/d}$. Starting from the Hamiltonian

$$H = \sum_i \frac{P_i^2}{2m} + \sum_{ij}^{N} x_i D_{ij} x_j \qquad (C1)$$

we will be interested in the fluctuations of the quantity $< E|x^2{}_o|E >$ as the energy eigenstate $|E >$ is varied. Here $x_i$ represents the displacement of the ith atom for equilibrium, and $D_{ij}$ is some arbitrary coupling between atoms. A good measure of this is $\Delta x^2 / << E|x^2|E >>_{micro}$ where

$$(\Delta x^2)^2 \equiv << E|x^2|E >^2>_{micro} - << E|x^2|E >>^2_{micro} \qquad (C2)$$

The subscript 0 has been dropped on $x_o$ for notational simplicity.

There is a system of normal coordinates which decouples (C1) into separate systems of the form (1). The total energy in an eigenstate is quantized

$$E = \sum_\nu \Omega_\nu n_\nu \qquad (C3)$$

(here the energy is taken from relative to its ground state value). In terms of normal coordinates, the Hamiltonian becomes

$$H = \frac{1}{2} \sum_\nu p_\nu^2 + \Omega_\nu^2 \eta_\nu^2 \qquad (C4)$$

In general, a coordinate $x$ can be written as

$$x = \sum_\nu a_\nu \eta_\nu \qquad (C5)$$

and therefore

$$< E|x^2|E> = \sum_\nu a_\nu{}^2 < E|\eta_\nu{}^2|E> = \sum_\nu a_\nu{}^2 \frac{1}{\Omega_\nu{}^2} (n_\nu + \frac{1}{2}) \qquad (C6)$$

To evaluate $\Delta x^2$, we examine the general relation between the microcanonical average of a quantity A and its canonical average. By applying the method of steepest descent, one has to leading order in $\frac{1}{N}$

$$< A>_{micro} = \frac{\int d\lambda \, e^{\lambda E} e^{-\lambda F(\lambda)} < A>_\lambda}{\int d\lambda \, e^{\lambda E} e^{-\lambda F(\lambda)}} \qquad (C7)$$

$$= < A>_\beta + \frac{1}{2} \frac{1}{\partial^2 \lambda F/\partial \lambda^2} \frac{\partial^2 < A>_\lambda}{\partial \lambda^2} + \frac{1}{2} \frac{\partial^3 \lambda F/\partial \lambda^3}{\left(\partial^2 \lambda F/\partial \lambda^2\right)^2} \frac{\partial < A>_\lambda}{\partial \lambda}$$

where the partial derivatives are evaluated at the saddle point $\lambda = \beta$ in the usual way. Here $< \cdots >_\lambda$ denotes the canonical average taken at temperature $1/\lambda$. Applying this formula to the problem at hand, on has to leading order in $\dfrac{1}{N}$

$$(\Delta x^2)^2 = \,\,<< E|x^2|E>^2>_\beta - \,<< E|x^2|E>>^2_\beta + \frac{1}{2}\frac{1}{\partial^2\lambda F/\partial\lambda^2}\left(\frac{\partial << E|x^2|E>>_\lambda}{\partial\lambda}\right)^2$$

Since the free energy is $\propto N$, the last term is $\propto 1/N$ for d > 1, $N$ for d=1 (since for d=1, $<< E|x^2|E>>_\beta \propto N$. Since for $\mu \neq \nu$, $< n_\mu n_\nu >_\beta = \,< n_\nu >_\beta < n_\mu >_\beta$, the first two terms in the above expression give

$$<< E|x^2|E>^2>_\beta - \,<< E|x^2|E>>^2_\beta = \sum_\nu \frac{a_\nu^4}{\Omega_\nu^2}\left[< n_\nu^2 >_\beta - \,< n_\nu >^2_\beta\right]$$

For harmonic crystal in d dimensions $a_\nu = const \propto 1/\sqrt{N}$, and the above expression can be estimated by turning it into an integral over k vectors, using the dispersion relation for small k that $\Omega = const \times k$. There is a divergence for small k in $d \leq 4$ dimensions giving

$$<< E|x^2|E>^2>_\beta - \,<< E|x^2|E>>^2_\beta \,\propto\, \begin{cases} N^{\frac{1}{d}-1} & d < 4 \\ 1/N & d > 4 \end{cases}$$

Comparing the contributions from these two terms one has, to leading order in $1/N$

$$\frac{(\Delta x^2)^2}{<< E|x^2|E>>^2_{micro}} \,\propto\, \begin{cases} N^{\frac{1}{d}-1} & 1 < d \;\; 4 \\ 1/N & d = 1 \; or \; d > 4 \end{cases}$$

**References**

1. S. K. Ma, ''*Statistical Mechanics*'', World Scientific (1985) and references therein;

2. I.E. Farquhar, ''*Ergodic Theory in Statistical Mechanics*'', Interscience publishers (1964).

3. J. von Neumann, Z. Physik **57**, 30 (1929).

4. M. Fierz, Helv. Phys. Acta **28**, 705 (1955).

5. H. Ekstein, Phys. Rev. **107**, 333 (1957).

6. M.V. Berry, Proc. Roy. Soc. Lond. A **413** 183 (1987).

7. E.P. Wigner, Ann. Math. **67**, 325 (1958).

8. R.U. Haq, A. Pandey, and O. Bohigas, Phys. Rev. Lett. **48** 1086 (1982).

9. M. Feingold and A. Peres, Phys. Rev. A **34** 591 (1986).

10. M. Feingold, D.M. Leitner, and O. Piro, Phys. Rev. A **39** 6506 (1989).

11. L.D. Landau and E.M. Lifshitz ''*Statistical Physics*'', Pergamon Press, London-Paris (1958).

12. S.F. Edwards and M. Warner, J. Phys. A **13** 381 (1980).

13. R.P. Feynman, ''*Statistical Mechanics*'', W.A. Benjamin (1972).

14. A.O. Caldeira and A.J. Legget, Ann. Phys. (N.Y.) **149** 374 (1984); R.P. Feynman and Vernon, Ann. Phys (N.Y.) **24** 118 (1963).

15. G.C. Stey and G. Gusman, J. Phys. C. **6** 650 (1973).

Figure 1. A plot of the eigenvectors $c_{ij}$ as a function of j for several different values of i. The dark shaded region represents the actual values taken by the components of one eigenvector, which appear as a continuous density in the limit of a large energy spread. The black curve on the top represents the envelope of this eigenvector $\sqrt{\Lambda(i,j)}$. The lighter shaded region represents the same thing for a different eigenvector, and the region of overlap between these two eigenvectors is denoted by the dark diamond shape region. In this region of overlap these two eigenvectors must be orthogonal. This leads to an effective repulsion between them. Curves to the right represent $\sqrt{\Lambda(i,j)}$'s for several different values of i. In this limit these curves should be strongly overlapping and have a uniform spacing.

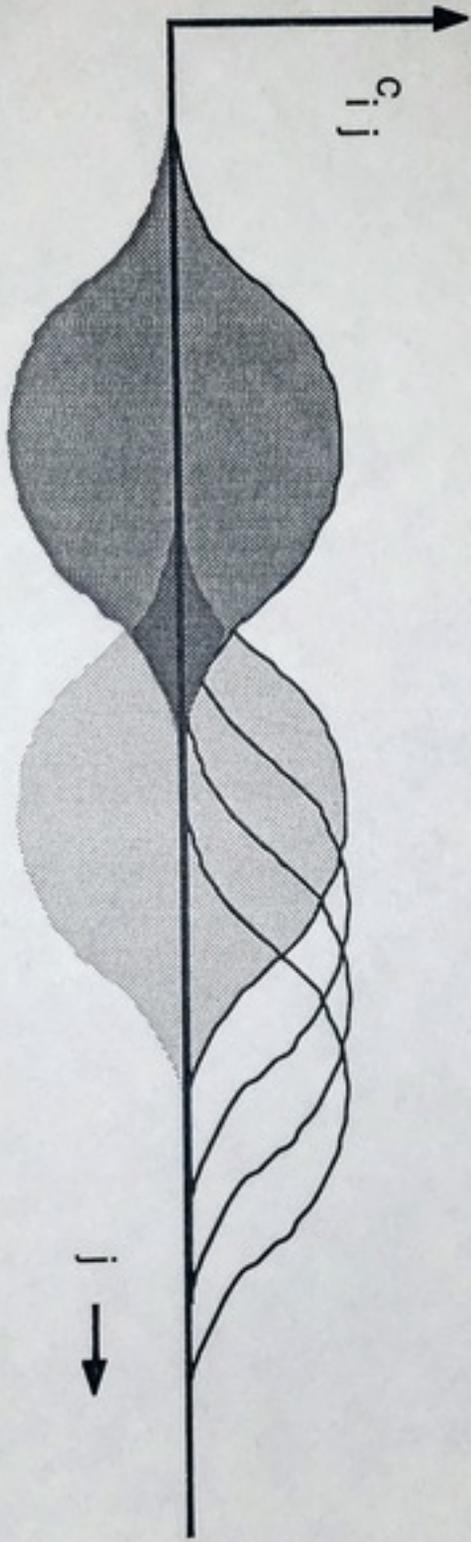